\documentstyle[aps,preprint]{revtex}

\newcommand{\cotH}{\mbox{$\cot \theta_H$}}
\newcommand{\thH}{\mbox{$\theta_H$}}
\newcommand{\thc}{\mbox{$\theta_c$}}
\newcommand{\thz}{\mbox{$\theta_0$}}
\newcommand{\epsn}{\mbox{$\epsilon$}}
\newcommand{\kkbar}{\mbox{$\bar{K}_2-\bar{K}_{2}^{'}$}}
\newcommand{\hc}{\mbox{$\bar{H}_c$}}
\newcommand{\ben}{\begin{equation}}
\newcommand{\een}{\end{equation}}
\newcommand{\delm}{\mbox{$\delta_m$}}
\newcommand{\ombbar}{\mbox{$\bar{\omega}_{b}$}}
\newcommand{\taubar}{\mbox{$\bar{\tau}$}}

\begin{document}
\draft
\title{Magnetic Field Dependence of Macroscopic
Quantum Tunneling and  Coherence of 
Ferromagnetic Particle}

\author{Gwang-Hee Kim$^a$ and Dae Sung Hwang$^b$}
\address{Department of Physics, Sejong University, Seoul 143-747
  , Korea}
\date{Received \hspace*{10pc}}
\maketitle

\begin{abstract}
  We calculate the quantum tunneling rate of a ferromagnetic
  particle of $\sim 100 \AA$ diameter
  in a magnetic field of arbitrary angle.
  We consider 
  the magnetocrystalline anisotropy with the biaxial symmetry
  and that with the 
  tetragonal symmetry.
  Using the spin-coherent-state path integral, we obtain
  approximate analytic formulas 
  of the tunneling rates in the small
  $\epsn(=1- H/H_c)$-limit 
  for the magnetic field 
  normal to the easy axis ($\thH = \pi/2$),
  for the field
  opposite to the initial easy axis ($\thH = \pi$),
  and for  
  the field at an angle
  between these two orientations
  ($\pi/2\ll\theta_H\ll\pi$). 
  In addition,
  we obtain numerically the
  tunneling
  rates for the biaxial symmetry 
  in the full range of the angle $\thH$ 
  of the magnetic field ($\pi/2 < \theta_H \leq \pi$),
  for the values of $\epsn=0.01$ and 0.001.\\
\end{abstract}             

\thispagestyle{empty}

\pacs{PACS numbers: 75.45.+j, 73.40.Gk, 75.30.Gw, 75.50.Gg\\
\\
\\
\\
\\
\\
\noindent
$a$ e-mail: gkim@phy.sejong.ac.kr\\
$b$ e-mail: dshwang@phy.sejong.ac.kr}

\section{Introduction}
\label{sec:intro}

A quest to understand the macorscopic 
quantum tunneling of 
ferromagnetic particles has
been an important 
topical issue of intensive theoretical\cite{chu,sch}
and experimental studies.\cite{aws}
Since 
the magnitude of the total magnetization
$\vec{M}$ is frozen up at sufficiently
low temperature, the direction $\hat{M}$ of the
total magnetization becomes
the only dynamical 
variable.
In the absence of an external
magnetic field, this direction is subject to
the magnetocrystalline anisotropy which possesses its
easy directions depending on the crystal symmetry.
In this situation, $\hat{M}$ has at least two equivalent
stable 
orientations
which leads to the oscillation of $\hat{M}$
between them.
This phenomenon is called macroscopic 
quantum coherence(MQC).\cite{chu}-\cite{aws}
Simple analysis shows that in this case the height of barrier is
too high to observe MQC in experiments. However, by applying a 
magnetic field $\vec{H}$ along the direction which lies halfway
between two equivalent 
stable orientations, we can control the height
and the width
of barrier and make MQC observable. 
On the other hand, if we apply a 
magnetic field in a proper direction, the symmetry of two
easy directions is broken. Then, one of the two 
stable orientations
becomes metastable. In this situation we can obtain the 
optimal condition for the observation of tunneling, which is called
macroscopic quantum tunneling(MQT).\cite{chu}-\cite{aws}
The ferromagnetic particle
is typically a single domain with as many as $10^5-10^6$ magnetic
moments, which is a macroscopic number of particles. During the 
dynamical process we can not neglect the influence of the
environment such as phonons,\cite{gar}
nuclear spins,\cite{gar1} and Stoner
excitations and eddy currents in metallic magnets.\cite{tat}
Even though some of these couplings are noticable, 
it has been reported\cite{gar}-\cite{tat} 
that they are not strong enough 
to make MQC or MQT unobservable. A few experimental attempts\cite{aws}
have
been made to observe the MQC or MQT 
of a large single ferromagnetic particle
or a collection of magnetic particles.
At
present, it is not easy to perform a direct comparison between
theoretical and experimental results because of stochastic
behaviors of the systems.

Recently, Zaslavskii\cite{zas} studied the 
uniaxial anisotropy which is
the simplest case, mapped the spin problem onto a 
one-dimensional particle system, and then obtained the 
tunneling exponent, the preexponential factor and
their temperature dependences in the low
barrier limit
with a magnetic field applied at some
angle to the anisotropy axis. Later, 
Miguel and Chudnovksy\cite{mig} performed
the calculation 
based on the imaginary time
path integral method, and 
demonstrated that their result
of the tunneling exponent in MQT coincides
with the 
Zaslavskii's result in the uniaxial symmetry,
for its dependence on the direction and 
magnitude of the applied magnetic field.
They discussed the 
tunneling rates
at finite temperature and suggested the experimental
procedures whose results can be compared with
their theorectical results.
In this paper, we extend the above
calculations to the
biaxial and tetragonal symmetries\cite{foot0}
by applying the instanton technique.
Since the result of the biaxial symmetry is a generalization of
that 
of the 
uniaxial symmetry 
studied by Zaslavskii, and by Miguel and Chudnovsky,
we can compare our results with theirs by taking
the 
vanishing limit of the transverse anisotropy constant.
Also, we will explain that MQC and MQT can be consecutively 
observed by changing the direction of $\vec{H}$, and discuss
their dependence on the direction and the magnitude of $\vec{H}$.

This paper is organized  as follows. In Sec. \ref{sec:mqt},
we briefly discuss the theory of MQT and MQC in a ferromagnetic
particle based on the standard instanton method. 
In Sec. \ref{sec:biax} and \ref{sec:tetr}, we
consider the tunneling rate for biaxial and tetragonal 
symmetry in a magnetic field with a range of angles
$\pi/2 \leq \thH \leq \pi$. We derive approximate 
formulas of the tunneling rates in the small $\epsn$-limit for 
three angle ranges 
($\thH=\pi/2, \ \pi/2 \ll \thH \ll \pi, \ \thH=\pi$),
and present the $\thH$-dependence of the
WKB exponent in the full range of 
angles ($\pi/2 < \thH \leq \pi$). The conclusions
are given in Sec. \ref{sec:conc}.

\section{The MQT and MQC in a Ferromagnetic System}
\label{sec:mqt}

The tunneling rate $\Gamma$
of a ferromagnetic 
particle escaping from a metastable
state in MQT or the oscillation rate 
$\Delta$ between double
wells in MQC has the relation

\begin{equation}
 (\Gamma \ \ \ {\rm or} \ \ \ \Delta)
 \ \ \ \propto
 \ \ \ \exp(-B).
\label{eq:rate1}
\end{equation}
The WKB exponent $B$ in Eq. (\ref{eq:rate1}) 
is approximately given by
$U/\hbar \omega_b$, where $U$ is the height of barrier.
The barrier frequency
$\omega_b$ is
the frequency of small oscillations around
the minimum of the inverted potential, and
characterizes the width of the barrier
hindering the decay process. 
$\omega_b^2$ is inversely proportional to the
effective
mass of the magnetic particle, 
where the mass is induced by the transverse
component of the magnetic field or by
the transverse magnetic anisotropy constant.
Therefore,
it is necessary that the magnetic particle has
a transverse magnetic anisotropy or 
we apply a transverse magnetic field,
in order that 
MQT or MQC of the magnetic particle is 
possible. 
For calculating $\Gamma$ or $\Delta$ in Eq. (\ref{eq:rate1}),
we introduce the angles $\theta$ and
$\phi$ for the direction of $\vec{M}$
in the spherical coordinate system and employ the instanton method,
in which the Euclidean action is written as
\begin{equation}
  S[\theta(\tau),\phi(\tau)]=V\int[i \frac{M_0}{\gamma}
  (1-\cos \theta) \frac{d \phi(\tau)}{d \tau} 
  + E(\theta(\tau),\phi(\tau))] d \tau , \label{eq:vis}
\end{equation}
where $V$ is the volume of the magnetic
particle, $\gamma=g \mu_B /\hbar$,
$\mu_B$ the Bohr magneton and $M_0$ the magnitude of the 
magnetization. The first term in Eq. (\ref{eq:vis}) 
is the topological
Wess-Zumino term\cite{los},
and the second term is the energy density which is composed of the
magnetic
anisotropy energy and the potential energy
given by external magnetic field.

The action (\ref{eq:vis})
produces from $\delta S=0$
the classical 
equation of motion
for $\vec{M}$ called the Gilbert equation\cite{gil},
which in the Euclidean space 
is written as
\begin{equation}
  i\frac{d \vec{M}}{d \tau} =- \gamma \vec{M} \times 
  \frac{dE}{d \vec{M}}.
\end{equation}
We note that the action (\ref{eq:vis}) describes the
$1 \oplus 1$ dimensional dynamics 
in the Hamiltonian formulation,
which consists of the canonical coordinates
$\phi$ and $p_{\phi}=1- \cos \theta$.\cite{joh,kim1}

In the standard instanton method, the 
tunneling or the oscillation rate is given by
the formula
\begin{equation}
\left \{
\begin{array}{c}
  \Gamma \\
  \Delta\\
\end{array}
\right \}
=
C_0 \sqrt {\frac{S_{cl}}{2 \pi \hbar}}
\left \{
\begin{array}{c}
  \omega_t \\
  \omega_c\\
\end{array}
\right \}
\exp(-\frac{S_{cl}}{\hbar}), \label{eq:rate}
\end{equation}
where $S_{cl}$ is  the classical 
action evaluated by using 
the solution of the equation of motion
derived from 
$\delta S=0$. In Eq. (\ref{eq:rate})
$C_0$ is the preexponential
factor 
which is originated 
from the $\delta^2 S$ term, $\omega_t$ a precession frequency
in MQT
and $\omega_c$ an oscillation
frequency in MQC, where 
$\omega_t$ and $\omega_c$ are of the order of the barrier
frequency $\omega_b$.\\

\section{Tunneling rate for biaxial symmetry}
\label{sec:biax}

In this section we study a system which has the 
biaxial symmetry, with $\pm \hat{z}$ axes 
being the easy axes of the Hamiltonian.
As in Ref. \cite{mig}, we apply the 
magnetic field in the $xz$-plane and take
the $\hat{z}$ axis as the initial 
easy axis when there is no magnetic field.
Then the total energy $E(\theta, \phi)$ of the 
system is given by
\begin{equation}
  E(\theta, \phi)=K_1 \sin^2 \theta + K_2 \sin^2 \phi
  \sin^2 \theta -H_x M_0 \sin \theta \cos \phi  -
  H_z M_0 \cos \theta + E_0 ,
\label{eq:toten}
\end{equation}                                                                 
where $K_1$ and $K_2$ are the parallel and transverse
anisotropy constants, respectively, and $E_0$ is
a constant which makes $E(\theta, \phi)$ zero at the initial
orientation. As will be shown later, the effective mass
of the magnetic particle is inversely 
proportional to a linear combination
of $K_2$ and transverse magnetic field $H_x$ 
while there is no exact analog of the kinetic energy
in the action (\ref{eq:vis}).
Denoting 
$\theta_{H}$ to be the angle between the magnetic
field and $z$-axis, we obtain MQC for $\theta_{H}= \pi /2$
and MQT for $\pi /2 < \theta_{H} \le \pi$ with large possible
tunneling rates depending on the magnitude 
of the applied magnetic
field.

It is convenient to introduce the dimensionless parameters
\begin{eqnarray}
 \bar{K}_2 \equiv K_2/2K_1, \ \ \ \bar{H}_x \equiv H_x/H_0, \ \ \
 \bar{H}_z \equiv H_z/H_0,
 \label{eq:dimpar}
\end{eqnarray}
where $H_0 \equiv 2K_1/M_0$. Then the total energy (\ref{eq:toten}) is
written as
\begin{equation}
 \bar{E}(\theta, \phi)=\frac{1}{2} \sin^2 \theta
 +\bar{K}_2 \sin^2 \phi \sin^2 \theta - \bar{H}_x
 \sin \theta \cos \phi  - \bar{H}_z \cos \theta + \bar{E}_0,
\label{eq:reden}
\end{equation}
where $\bar{E}(\theta, \phi)=E(\theta, \phi)/2K_1$.
The plane given by $\phi =0$ is the easy plane, on which 
$\bar{E}(\theta, \phi)$ is given by
\begin{equation}
 \bar{E}(\theta,0)=\frac{1}{2} \sin^2 \theta
 -\bar{H} \cos (\theta- \theta_H ) + \bar{E}_0,
\end{equation}
where $\bar{H}_x=\bar{H} \sin \theta_H$ and
$\bar{H}_z=\bar{H} \cos \theta_H$ since $\vec{H}$ is in $xz$-plane.
We define $\theta_0$ to be the initial angle and 
$\theta_c$ the angle at which the barrier vanishes by
the applied critical magnetic field.
From Eq. (\ref{eq:reden}) we get the following conditions
which
$\theta_0$ and $\theta_c$ should satisfy,
\begin{eqnarray}
 \sin \theta_0 \cos \theta_0 + \bar{H} \sin 
 ( \theta_0 - \theta_H)=0, \label{eq:th0} \\
 \sin \theta_c \cos \theta_c + \bar{H}_c \sin
 ( \theta_c - \theta_H)=0, \label{eq:thc} \\
 \cos(2 \theta_c) + \bar{H}_c \cos
 ( \theta_c - \theta_H)=0, \label{eq:thc1}
\end{eqnarray}
where $\bar{H}_c$ is the dimensionless critical field
with which the barrier just vanishes.
From Eqs. (\ref{eq:thc}) and 
(\ref{eq:thc1}) the critical field $\bar{H}_c$
is given by
\begin{equation}
 \bar{H}_c =\frac{1}{(\sin ^{2/3} \theta_H +
 |\cos \theta_H|^{2/3})^{3/2}}.
\label{eq:hcb}
\end{equation}
From Eqs. (\ref{eq:thc})-(\ref{eq:hcb}) 
$\theta_c$ is given by 
\begin{eqnarray}
 \sin(2\thc)&=&{2|\cot\thH|^{1/3}\over 
 1+|\cot\thH|^{2/3}}, \label{eq:sthcb}\\
 \cos(2\thc)&=&1-{2\over 1+|\cot\thH|^{2/3}}.
 \label{eq:cthcb}
\end{eqnarray}
For the special cases
$\theta_H= \pi/2, \ 3\pi/4, \ \pi$, we have 
$\theta_c = \pi/2, \  \pi/4, \  0$,
respectively.

The practically interesting situation is when 
the barrier height is low and 
the width is narrow in order to have the
tunneling rate large.
Such a situation is realized when
the value of 
$\epsilon \equiv 1-\bar{H}/\bar{H}_c$ is small.
For the small value of $\epsn$,
Eq. (\ref{eq:th0}) becomes
\begin{equation}
\sin(2 \theta_c) (\epsilon - \frac{3}{2} \eta^2)
- \eta \cos(2 \theta_c)(2 \epsilon - \eta^2)=0,
\label{eq:etaeq}
\end{equation}
where $\eta=\theta_c - \theta_0$ which is 
expected to be small for $\epsn \ll 1$. 
As will be seen later, detailed calculations
show that $\eta$ is of the order of $\sqrt{\epsn}$.
Then
the order of magnitude of the second term in
Eq. (\ref{eq:etaeq}) is smaller than that
of the first term by $\sqrt{\epsn}$ and the value of
$\eta$ is determined by the first term to have
$\eta \simeq \sqrt{2 \epsilon /3}$,
except for the $\thH$ values
near $\theta_H=\pi /2$ and $\pi$. 
However, when $\thH$ is very close to 
$\pi/2$ or $\pi$, $\sin (2 \theta_c)$
becomes almost zero as shown in Fig.~1,
and the first term is much smaller than the
second term in Eq. (\ref{eq:etaeq}).
Then 
the value of $\eta$ is obtained from
the second term 
when $\thH \simeq \pi/2$ or $\pi$, and
$\eta$ is given by
$\eta \simeq \sqrt{2 \epsilon}$
for $\theta_H \simeq \pi/2$ and $\eta \simeq 0$
for $\theta_H \simeq \pi$ 
from the detailed analysis of 
Eqs. (\ref{eq:th0})-(\ref{eq:thc1}).
After a little manipulation, we obtain from 
Eq. (\ref{eq:reden})
an approximate formula
of $\bar{E} (\theta, \phi)$ in the limit of small $\epsilon$
given by
\begin{equation}
 \bar{E}(\delta, \phi)=\bar{K}_2 \sin^2 \phi \sin^2 
 (\theta_0 + \delta) + \bar{H}_x \sin (\theta_0+\delta)
 (1-\cos \phi) + \bar{E}_1(\delta),
\label{eq:totdel}
\end{equation}
where $\bar{E} (\theta, \phi)$ is written as
$\bar{E}(\delta, \phi)$ by introducing 
a small variable $\delta \equiv
\theta - \theta_0$, and  $\bar{E}_1(\delta)$ is a function of 
only $\delta$ given by
\begin{equation}
 \bar{E}_1 (\delta)=\frac{1}{4} \sin (2 \theta_c)(3 \delta^2
                 \eta - \delta^3)+\frac{1}{2} \cos
	         (2 \theta_c)[\delta^2 (\epsilon -
	         \frac{3}{2} \eta^2)+\delta^3 \eta - 
	         \frac{\delta^4}{4}].
\label{eq:edelta}
\end{equation}
As previously discussed for $\eta$, even though the 
$\cos(2\thc)$ term
in Eq. (\ref{eq:edelta}) looks smaller by a factor of 
$\eta$ which is 
of the order of $\sqrt{\epsilon}$, the second term can not be neglected
near $\theta_H = \pi/2$ and $\pi$ because $\sin ( 2 \theta_c)$
is almost zero for these regions of $\thH$.
When we assume that 
$|\phi | \ll 1$, from the energy 
conservation $\bar{E}(\delta,\phi)=0$ in 
Eq. (\ref{eq:totdel}),
$\phi$ is approximately given by
\begin{equation}
 \phi^2 = -\frac{\bar{E}_1(\delta)}{
         \bar{K}_2 \sin^2(\theta_0+\delta)+\frac{1}{2}
	 \bar{H}_x \sin(\theta_0 + \delta)}.
\label{eq:ephi}
\end{equation}
Since $\delta$ is of the order of $\sqrt{\epsn}$ as can be
seen from Eq. (\ref{eq:edelta}), $\bar{E}_1(\delta)$ is 
$O(\epsn^{3/2})$ or less. Therefore, $\phi$ is very
small from Eq. (\ref{eq:ephi}) whose validity
is for the full range of angles 
$\pi/2 \leq \thH \leq \pi$ in biaxial symmetry.
However, in case of tetragonal symmetry the magnitude of 
$\phi$ is not small about $\thH=\pi$, as can be seen in
Eq. (\ref{eq:phi180t}). 
In such a situation it is not possible 
to expand $\bar{E}(\delta,\phi)$ as powers of 
$\phi$ and to reduce the classical equation
to the one-dimensional equation like Eq. (\ref{eq:deltaeq}).

Since we confirmed that $|\phi|$ is very small
in case of the biaxial symmetry which we study
in this section, we have the approximate formula
of $\bar{E}(\delta, \phi)$ in Eq. (\ref{eq:totdel})
given by
\begin{equation}
 \bar{E}(\delta,\phi) = [\bar{K}_2 \sin^2 (\theta_0 +
 \delta) + \frac{1}{2} \bar{H}_x \sin (\theta_0 + \delta)]
 \phi^2 + \bar{E}_1(\delta). \\
\label{eq:int}
\end{equation}
By introducing a new 
scaled time variable $\bar{\tau} \equiv \omega_0 \tau$
with $\omega_0 \equiv 2 \gamma K_1 /M_0$, the Euclidean
action (\ref{eq:vis}) becomes
\begin{equation}
  S[\delta(\taubar),\phi(\taubar)]= \hbar J
  \int \{ i [1-\cos (\theta_0 + \delta)
  \frac{d \phi(\taubar)}{d \bar{\tau}}]
  + \bar{E}(\delta(\taubar),\phi(\taubar))\} d \bar{\tau} , 
\label{eq:newactn}
\end{equation}
where $J \equiv V M_0 / \hbar \gamma$. 
In the following we also use the scaled 
angular frequencies $\bar{\omega}_i$ defined by
$\bar{\omega}_i \equiv \omega_i / \omega_0$
($i=b,\  c, \  t$). 

There are three approaches for the calculation
of $S_{cl}$ from the action (\ref{eq:newactn}). Firstly, if 
$|\phi|$ is small, 
we can perform a
Gaussian
integration over the variable $\phi$
in the path integral
and reduce the system to that
with only one variable $\delta$. Then it is
possible to do the rest of the calculation by using
the standard instanton method.\cite{mig}
However, if $|\phi|$ is not small like 
in the tetragonal symmetry case
with $\thH=\pi$, the integrand
of Eq. (\ref{eq:newactn}) can not be reduced to the form
which allows the Gaussian integration over 
$\phi$.
Secondly, we can use the fact that $\bar{E}(\delta_{cl},
\phi_{cl})=0$ and that $(d \phi/d \delta)d \delta$
can be substituted for $(d \phi/d \bar{\tau})d \bar{\tau}$
in the Euclidean action (\ref{eq:newactn}). 
Here, we do not need to know the 
explicit solutions of the classical paths 
$\delta_{cl}(\taubar)$
and $\phi_{cl}(\taubar)$ for the classical action $S_{cl}$
because 
we can obtain
$d \phi/d \delta = (d \phi/d \bar{\tau})/
(d \delta /d \bar{\tau})$ from 
the Euler-Lagrange equations given by
Eq. (\ref{eq:el1}) and (\ref{eq:el2}) below.
However, this approach can not give the value
of the preexponential factor, for which we need the 
explicit solution
of the bounce in MQC or the instanton in MQT 
as a function of $\taubar$.\cite{col}
Thirdly, we directly solve the coupled equations 
of motion for 
$\delta$ and $\phi$ given by 
Eqs. (\ref{eq:el1}) and (\ref{eq:el2})
by incorporating $\bar{E}(\delta_{cl}, \phi_{cl})=0$.
Even though it is hard and sometimes rather tedious to obtain
the solution as a function of $\taubar$ from Eqs.
(\ref{eq:el1}) and (\ref{eq:el2}), and 
$\bar{E}(\delta_{cl}, \phi_{cl})=0$, this methods
provides a complete result for the tunneling rate $\Gamma$ in MQT
or the oscillation frequency $\Delta$ in MQC.
In
this approach, the preexponential factor
can be obtained from 
the second variational term $\delta^2 S$ of the action
in the spin-coherent-path integral,
and
the classical action $S_{cl}$ is obtained 
by the direct integration
of Eq. (\ref{eq:newactn}) over $\taubar$ 
with the explicit solution 
$\delta_{cl}(\taubar)$ and $\phi_{cl}(\taubar)$. 
While the third method is the most general approach
to obtain the tunneling rate, it is reduced to the 
first one if $|\phi|$ is small enough. 
Also, the second method is
useful for checking
the result of 
the first or the third method for the WKB exponent.
In 
biaxial symmetry, due to $|\phi| \ll 1$
we apply the first method
to obtain
the explicit instanton or bounce 
solution as a function of
$\taubar$, which is used to calculate
the WKB exponent and the
preexponential factor. On the other hand,
since the magnitude of $\phi$ is not
small in tetragonal symmetry case with $\thH=\pi$,
we need to make use of the third method 
for the calculation of $\Gamma$.
In addition, by using the second method we check
the results of the WKB exponent from the
first or third method
for each case.

The classical 
trajectory of 
$\delta$ and $\phi$ is determined by the 
Euler-Lagrange equation derived from the action
(\ref{eq:newactn}),
\begin{eqnarray}
 i \sin(\theta_0 + \delta) \frac{d \phi}{d \bar{\tau}}
         &=& - \frac{\partial \bar{E}(\delta, \phi)}
           {\partial \delta},  \label{eq:el1} \\
 i \sin(\theta_0 + \delta) \frac{d \delta}{d \bar{\tau}}
         &=&  \frac{\partial \bar{E}(\delta, \phi)}
           {\partial \phi}.       \label{eq:el2}
\end{eqnarray}
From Eqs.
(\ref{eq:int}), (\ref{eq:el1}) and (\ref{eq:el2}), the equation
of motion for the instanton $\delta_{cl}(\taubar)$ in the
biaxial symmetry case becomes
\begin{equation}
 \frac{d \delta_{cl}}{d \bar{\tau}} =
  \sqrt{\frac{2}{M} \bar{E}_1(\delta_{cl})}  \  ,
\label{eq:deltaeq}
\end{equation}
where the effective mass is given by
\begin{equation}
 M=[\frac{\bar{H}_x}{\sin(\theta_0+\delta_{cl})}+2 \bar{K}_2]^{-1}.
 \label{eq:effmass90b}
\end{equation}
Now let us study  the tunneling rates
at three different magnetic field directions of
$\theta_H = \pi /2$, $\pi$, and 
$\pi/2 \ll \thH \ll \pi$.\\ 

\subsection{$\thH=\pi/2$}

For $\thH=\pi/2$, $\theta_c = \pi /2$ from 
Eqs. (\ref{eq:sthcb}) and (\ref{eq:cthcb}),
and $\eta = \sqrt{2 \epsilon}$ as explained 
below Eq. (\ref{eq:etaeq}).
Then, from 
Eqs. (\ref{eq:edelta}) and 
(\ref{eq:effmass90b}) we get 
the approximate forms of the
effective potential energy 
$\bar{E}_1(\delta)$ and the effective mass $M$ in 
the reduced one dimension as 
\begin{eqnarray}
 \bar{E}_1(\delta) &=& \frac{1}{2} \delta^2
 (\frac{\delta}{2}-\sqrt{2 \epsilon})^2, 
 \label{eq:en90b} \\
 M &=& (\bar{H}_x + 2 \bar{K}_2)^{-1},
 \label{eq:m90b}
\end{eqnarray}
where $\bar{H}_x \approx \bar{H}_c$. Here
we notice that MQC can not be generated without an
external field along the $x$-axis or a transverse 
anisotropy constant $K_2$.
Eq. (\ref{eq:deltaeq}) with Eqs. (\ref{eq:en90b})
and (\ref{eq:m90b}) has the solution 
called the bounce which is given by 
\begin{equation}
 \delta_{cl}(\taubar) = \sqrt{2 \epsilon} 
 [ 1+\tanh(\bar{\omega}_c \bar{\tau})],
\label{eq:delcl}
\end{equation}
where $\bar{\omega}_c = \sqrt{\epsn/2}\sqrt{1+K_2/K_1}$.
Then from Eq. (\ref{eq:ephi}) we get 
the classical path for $\phi$ given by
\begin{equation}
 \phi_{cl}(\taubar) = -i \epsilon
 {1\over \sqrt{1+{K_2 \over  K_1}}}
 [\frac{1}{\cosh^2(\bar{\omega}_c \bar{\tau})}] .
\end{equation}
We can calculate the 
classical action by integrating Eq. (\ref{eq:newactn})
with the above classical trajectory, and the result
is given by
\begin{equation}
S_{cl}=(\hbar J) \frac{4 \sqrt{2}}{3} \frac{\epsilon^{3/2}}
 {\sqrt{1+{K_2 \over K_1}}}.
\label{eq:sclbi90}
\end{equation}
From Eq. (\ref{eq:en90b}) and Fig.~2, we get 
the height of barrier as 
$\bar{E}_1(=U/2K_1V)=\epsilon^2/2$ at $\delta_m=\sqrt{2\epsn}$ and
the oscillation frequency around the minimum of the inverted
potential $-\bar{E}_1(\delta)$ as 
$\bar{\omega}_b(\equiv
\sqrt{-\bar{E}_{1}^{''}(\delta_m)/M})
=\sqrt{2} \, \bar{\omega}_c$.
Then,
as mentioned in Sec. \ref{sec:mqt}, 
the WKB exponent $B(=S_{cl}/\hbar)$
is approximately given by
\begin{equation}
 B \sim {U \over \hbar \omega_b}= 
 \frac{J}{2} \frac{\epsilon^{3/2}}{\sqrt{1+\frac{K_2}{K_1}}},
\end{equation}
which agrees 
up to the numerical factor with the result in Eq. (\ref{eq:sclbi90}) 
obtained by using the explicit instanton solution.\\

\subsection{$\pi/2 \ll \thH \ll \pi$}

For $\pi/2 \ll \thH \ll \pi$, the critical angle 
$\thc$ is in the range of
$0 \ll \theta_c \ll \pi/2$ and $\eta \simeq \sqrt{2 \epsilon /3}$.
Then 
from
Eqs. (\ref{eq:sthcb}) and (\ref{eq:cthcb})
we get
\begin{eqnarray}
 \bar{E}_1(\delta) &=& \frac{1}{4} \sin(2 \theta_c)
 (\sqrt{6 \epsilon} \delta^2 - \delta^3),
 \label{eq:en135b}\\
 M &=& [\frac{\bar{H}_x}{\sin \theta_c} + 2 \bar{K}_2]^{-1},
\end{eqnarray}
where $\bar{H}_x=\bar{H}_c \sin\thH$.
The classical equation of motion 
Eq. (\ref{eq:deltaeq})
gives the instanton solution as
\begin{eqnarray}
 \delta_{cl}(\bar{\tau}) &=& \frac{\sqrt{6 \epsilon}}
 {\cosh^2(\bar{\omega}_t \bar{\tau})}, \label{eq:delcl135}\\
 \phi_{cl}(\bar{\tau}) &=& -i (6\epsilon)^{3/4}
 |\cot\thH|^{1/6}
 \frac{\sinh(\bar{\omega}_t \bar{\tau})}
 {\cosh^3(\bar{\omega}_t \bar{\tau})}
 [1+({K_2\over K_1}){1\over {1+|\cot\thH|^{2/3}}}]^{-1/2},
\end{eqnarray}
where 
Eq. (\ref{eq:ephi}) has been used to get 
$\phi_{cl}(\bar{\tau})$, and 
the dimensionless frequency $\bar{\omega}_t$ is 
given by
\begin{equation}
\bar{\omega}_t=
(\frac{3}{8})^{1/4}\epsilon^{1/4}
\frac{|\cotH|^{1/6}}
{1+|\cotH|^{2/3}}
\sqrt{1+\frac{K_2}{K_1}(1+|\cotH|^{2/3})}.
\end{equation}
Then the
classical action is found to be
\begin{equation}
S_{cl}= (\hbar J) \frac{16 \times 6^{1/4}}{5}
\epsilon^{5/4} \frac{|\cot \theta_H|^{1/6}}
{\sqrt{1+\frac{K_2}{K_1}(1+|\cot \theta_H|^{2/3})}}.
\label{eq:scl135}
\end{equation}
It is noted that Eq. (\ref{eq:scl135}) without $K_2$
agrees with the classical action
in Ref. \cite{mig} which studied the uniaxial symmetry
case.
By using 
$\bar{E}_1(\delta_m=2\sqrt{6\epsilon}/3)
=\sin(2\theta_c)(6\epsilon)^{3/2}/27$ and 
$\bar{\omega}_b=\sqrt{-\bar{E}_{1}^{''}
(\delta_m)/M}=2\bar{\omega}_t$
given from Eq. (\ref{eq:en135b}) and 
Fig.~2, we approximately
obtain $B$ as
\begin{equation}
\frac{U}{\hbar \omega_b}=\frac{4\times 6^{1/4}}{9}
J \epsilon^{5/4}\frac{|\cot \theta_H|^{1/6}}
{\sqrt{1+\frac{K_2}{K_1}(1+|\cot \theta_H|^{2/3})}},
\end{equation}
which is consistent with Eq. (\ref{eq:scl135})
up to the numerical factor.\\

\subsection{$\thH=\pi$}

In case of $\theta_H =\pi$, 
we had $\theta_c=0$ and  $\eta=0$. 
Then we have 
\begin{eqnarray}
\bar{E}_1(\delta) &=& \frac{1}{2} (\epsilon
\delta^2 - \frac{\delta^4}{4}),\\
M &=& (2 \bar{K}_2)^{-1}
\end{eqnarray}
which gives the classical path obtained from
Eq. (\ref{eq:deltaeq}) as
\begin{eqnarray}
 \delta_{cl}(\bar{\tau})&=& \frac{2 \sqrt{\epsilon}}
 {\cosh(\bar{\omega}_t \bar{\tau})}, \label{eq:delcl180}\\
 \phi_{cl}(\bar{\tau})&=& -i \sqrt{K_1 \epsn \over K_2}
 \tanh(\bar{\omega}_t \bar{\tau}) + n \pi, 
\end{eqnarray}
where $n=0$ or $1$, and 
$\bar{\omega}_t=\sqrt{K_2 \epsn/K_1}$.
Here we note that $\phi_{cl}(\taubar)$ is obtained from the
conservation of energy
$\bar{E}(\delta_{cl}, \phi_{cl})=0$.
Then the corresponding classical action is
given by
\begin{equation}
 S_{cl}=(\hbar J)
 \frac{8}{3} \sqrt{\frac{K_1}{K_2}} \epsilon^{3/2}.
\label{eq:scl180bi}
\end{equation}
We note again
that
\begin{equation}
 B \sim  
 \frac{U}{\hbar \omega_b}= 
 \frac{J}{4}\sqrt{\frac{K_1}{K_2}}
 \epsilon^{3/2}, 
\end{equation}
which is obtained
from the fact that $\bar{E}(\delta_m=\sqrt{2\epsn})
=\epsilon^2/2$ and
$\bar{\omega}_b=\sqrt{2}\bar{\omega}_t$,
is consistent with $S_{cl}$ in 
Eq. (\ref{eq:scl180bi}) up to the numerical factor.\\

\subsection{Prefactor and Discussion}

In order to complete our study of 
MQT and MQC in the biaxial symmetry,
we need to calculate the preexponential 
factor in Eq. (\ref{eq:rate}).
Since in the biaxial symmetry the problem can be 
reduced to the one-dimensional one due to the smallness
of $|\phi|$, the calculation of the preexponential factor can be
performed by using the well-known instanton method developed by 
many authors.\cite{col}-\cite{col1} However, 
in many other 
symmetries it is not possible to reduce the problem
to the one-dimensional one by directly integrating over
one of two variables such as $\phi$, because the magnitude
of $\phi$ is not 
small enough for the action to be expaneded
as powers of $\phi$ for Gaussian integration. This
situation will be seen in the tetragonal symmetry case with 
$\theta_H=\pi$. In order to treat such a case, 
Garg and Kim\cite{kim}
studied the formalism for evaluation of the prefactor based on
the spin-coherent-state path integral. Here we explain briefly
the basic idea of this study. Expanding 
the action $S[\theta(\tau),\phi(\tau)]$ 
in terms of $\theta_1$ and $\phi_1$
about the classical path, 
where $\theta(\tau)=
\theta_{cl}+ \theta_1$ and $\phi(\tau)=
\phi_{cl}+ \phi_1$, we obtain the action as
$S[\theta,\phi] \simeq S_{cl}+\delta^2 S$ with
$\delta^2 S$ being a functional of $\theta_1$
and $\phi_1$. Then Gaussian integration can be
performed over $\phi_1$, 
and the $\theta_1$ path integral
is now cast into the standard form of
one-dimensional potential problem. What we need in this
calculation is the approximate form of $d\theta_{cl}/d\tau$ 
for large $\tau$. 

In the biaxial symmetry,
by using Eq. (\ref{eq:delcl}) for 
$\thH=\pi/2$, 
Eq. (\ref{eq:delcl135}) for $\pi/2 \ll \theta_H \ll \pi$,
and Eq. (\ref{eq:delcl180}) for $\thH=\pi$,
we obtain the complete anlaytic forms of the
tunneling rate for MQT or the oscillation rate for MQC, 
and the results are
summarized in Table~II(a).\cite{com,cal} We note that in Table~II(a)
the preexponential factor
is multiplied by
a factor of two 
in case of $\thH=\pi$ because 
there exist two 
classical paths which give the same action.
Also it is noted that the $\epsilon$ behavior of the
WKB exponent $B$ is given by $\epsn^{3/2}$ for $\thH=\pi/2$,
$\epsn^{5/4}$ for $\pi/2 \ll \thH \ll \pi$, and
$\epsn^{3/2}$ for $\thH=\pi$.
It is seen 
by taking $K_2$ to be zero in Table~II\cite{foot1}
that the tunneling rate in the uniaxial symmetry
vanishes for $\thH=\pi$.  
This situation can be
understood from the fact that 
for $\thH=\pi$
the Possion bracket 
$\{p_{\phi}, H\}$ which determines the dynamics 
of the spin system with the Lagrangian
(\ref{eq:vis}) is zero 
because in this case
the Hamiltonian $H$
becomes a function of only $p_{\phi}$,
which gives 
$dp_\phi/d\tau =d(1-\cos\theta)/d\tau=0$
and then $\theta$ is constant in time.\cite{kim1}

We obtained the instanton solutions for the 
full range of $\thH$
($\pi/2 < \thH \le \pi$) by solving numerically 
the equations of motion (\ref{eq:el1}) and (\ref{eq:el2})
or equivalently Eq.  (\ref{eq:deltaeq}). Then
by using the obtained instanton solutions
we calculated the classical action from Eq. (\ref{eq:newactn})
to obtain the
WKB exponent $B(=S_{cl}/\hbar)$.
In Fig.~3 we present the instanton solutions with
$\epsn=0.001$ and $K_1=K_2$ for several values of
$\thH$ which we obtained by numerical calculations.
We also obtained the $\thH$-dependence of $B$
with $\epsn=0.01$ and $\epsn=0.001$ for
$\pi/2 < \thH \le \pi$, and present the result
in Fig.~4. As is noted in the figure, the maximal value of
$B$ is at about $\thH=2.78(\simeq 159^{\circ})$ and the approximate
analytic
result obtained 
in Eq. (\ref{eq:scl135}) is almost valid in the range of
angles $120^{\circ} \leq \thH \leq 170^{\circ}$.\\

\section{Tunneling rate for tetragonal symmetry}
\label{sec:tetr}

In this section we study the 
tetragonal symmetry whose anisotropy
energy is given by
\begin{equation}
E_a (\theta,\phi)=K_1 \sin^2 \theta+K_2 \sin^4 \theta
  -K_{2}^{'} \cos(4\phi) \sin^4 \theta,
\end{equation}
where we once again take the easy axis to be $\pm \hat{z}$
for $K_1 > 0$. When we apply $\vec{H}$ in the 
$xz$-plane as in the previous section,
the total energy becomes
\begin{equation}
E(\theta,\phi)=E_a(\theta,\phi)-H_x M_0 \sin \theta \cos \phi  -
  H_z M_0 \cos \theta + E_0 ,
\end{equation}
where we assume that $K_1>0$ and $|\kkbar| \ll 1$.\cite{foot2}
Here
we also use the dimensionless parameters 
defined in Eq. 
(\ref{eq:dimpar}).
By choosing $K_{2}^{'}>0$, we take $\phi=0$ to 
be an easy plane in our calculations. 
In the 
$\phi=0$ plane the scaled total energy is written as
\begin{equation}
 \bar{E}(\theta,0)=\frac{1}{2} \sin^2 \theta +
 (\bar{K}_2-\bar{K}_{2}^{'}) \sin^4 \theta
 -\bar{H} \cos (\theta- \theta_H ) + \bar{E}_0,
\label{eq:entethe}
\end{equation}
where $\bar{K}_{2}^{'} \equiv  K_{2}^{'}/2K_1$.
The initial angle $\theta_0$ and the critical angle
$\thc$ defined in the previous section
are determined by the equations
\begin{eqnarray}
 \sin \theta_0 \cos \theta_0 + \bar{H} \sin
 ( \theta_0 - \theta_H)
 +4(\kkbar)\sin^3 \thz \cos \thz =0, \label{eq:th0te} \\
 \sin \theta_c \cos \theta_c + \bar{H}_c \sin
 ( \theta_c - \theta_H)
 +4(\kkbar)\sin^3\thc\cos\thc=0, \label{eq:thcte} \\
 \cos(2 \theta_c) + \bar{H}_c \cos
 ( \theta_c - \theta_H)
 +4(\kkbar)(3\sin^2\thc\cos^2\thc-
 \sin^4\thc)=0. \label{eq:thc1te}
\end{eqnarray}
Using Eqs. (\ref{eq:thcte}) and (\ref{eq:thc1te}),
we obtain the critical magnetic field as\cite{mar}
\begin{equation}
 \bar{H}_c =\frac{1}{(\sin ^{2/3} \theta_H +
 |\cos \theta_H|^{2/3})^{3/2}}
 [1+\frac{4(\kkbar)}{1+|\cos \theta_H|^{2/3}}].
\end{equation}
In the small $\epsn$-limit 
$(\epsn \equiv 1-\bar{H}/\bar{H}_c)$,
by using Eqs. (\ref{eq:thcte}) and (\ref{eq:thc1te})
we obtain the approximate equation for 
$\eta( \equiv \thc-\thz)$ given by
\begin{eqnarray}
 &-&\epsn[2\hc\sin(\thc-\thH)]
 +\eta^2[3\hc\sin(\thc-\thH)+6(\kkbar)\sin(4\thc)] 
 \nonumber \\
 & &+\eta\{\epsn [2\hc\cos(\thc-\thH)]
 -\eta^2[\hc\cos(\thc-\thH)+8(\kkbar)\cos(4\thc)]\}=0.
\label{eq:etaeqte}
\end{eqnarray}

The order of magnitude of the first two terms
in Eq. (\ref{eq:etaeqte}) is higher
than that of the
last third term by $O(\sqrt{\epsn})$.
Therefore, for 
$\pi/2 \ll \thH \ll \pi$ we obtain $\eta$ from
the first two terms of Eq. (\ref{eq:etaeqte})
as 
\begin{equation}
 \eta= \sqrt{{2\epsn}\over 3}[1+4 (\kkbar) \cos(2\thc)],
\end{equation}
and $\thc$ from Eqs. (\ref{eq:thcte}) and (\ref{eq:thc1te})
as 
\begin{equation}
\thc={\pi \over 4} +{4(\kkbar) \over 3}.
\end{equation}
However, for $\thH=\pi/2$ and $\pi$, since
$\sin(\thc-\thH)$ and $\sin(4\thc)$ are equal to zero,
the first two terms in Eq. (\ref{eq:etaeqte})
vanish. In these cases the value of
$\eta$ is determined by the last third term, which gives
\begin{eqnarray}
 \thc = {\pi\over 2}, \ \ \
 \eta \simeq \sqrt{2\epsn}[1-4(\kkbar)] \ \ \
 {\rm for}\ \ \thH &=& {\pi\over 2},\\
 \thc = 0, \ \ \
 \eta=0 \ \ \ \ \ \ \ \ \ \ \ \ \ \ \ \ \ \ \ \ \ \ \ \ \ \ \ \ \
 {\rm for} \ \ \thH &=& {\pi}.
\end{eqnarray}
For reference, we note that 
$\eta$ is approximately given by
\begin{equation}
 \eta \simeq \sqrt{2\epsn}[1-4(\kkbar) {\cos(4\thc)\over \bar{H}_c
 \cos(\thc-\thH)}]
\end{equation}
for the vaule of $\thH$ around $\thH=\pi/2$. 

In the small $\epsn$-limit the approximate form of 
$\bar{E}(\theta,0)$ in Eq. (\ref{eq:entethe})
is written as 
\begin{eqnarray}
\bar{E}_1(\delta)= &-&{1\over 2}[\hc \sin(\thc-\thH)+2(\kkbar) 
\sin(4\thc)](3 \delta^2 \eta- \delta^3) \nonumber \\
&-&{1\over 2}[\hc \cos(\thc-\thH)+8(\kkbar)
\cos(4\thc)][\delta^2(\epsn-{3\over 2} \eta^2)
+\eta \delta^3 -{\delta^4\over 4}] \nonumber \\
&+&4(\kkbar)\cos(4\thc)\delta^2 \epsn,
\end{eqnarray}
where $\delta \equiv \theta-\thz$ which is small 
in the small 
$\epsn$-limit. Then the total energy becomes
\begin{equation}
\bar{E}(\delta,\phi)=\bar{K}^{'}_2 [1-\cos(4\phi)]\sin^4(\thz+\delta)
+\bar{H}_x (1-\cos \phi) \sin(\thz+\delta) + \bar{E}_1(\delta),
\end{equation}
which will be discussed for three angle ranges, $\thH=\pi/2,\ \pi$
and $\pi/2 \ll \thH \ll \pi$.\\

\subsection{$\thH=\pi/2$}

Using the equation of motion for the classical trajectories,
Eqs. (\ref{eq:el1}) and (\ref{eq:el2}), the classical path
at $\thH=\pi/2$ is given by
\begin{eqnarray}
 \delta_{cl}(\bar{\tau})&=& \sqrt{2\epsn}
 [1-2{{(K_2-K_{2}^{'})}\over K_1}][1+\tanh(\bar{\omega}_c \bar{\tau})],
 \label{eq:del90t}\\
 \phi_{cl}(\bar{\tau})&=& -i \epsn
 {[1-2{{(K_2+K_1)}\over K_1}]}[{1\over \cosh^2(\bar{\omega}_c
 \bar{\tau})}],
 \label{eq:phi90t}
\end{eqnarray}
where $\bar{\omega}_c = \sqrt{\epsn/2}[1+2(K_2+K_{2}^{'})/{K_1}]$.
In this case
the approximate form of
$\bar{E}_1(\delta)$
and the effective mass become
\begin{eqnarray}
\bar{E}_1(\delta)&=&{[1+12(\kkbar)]\over 8}\delta^2
\{\delta-2\sqrt{2\epsn}[1-4(\kkbar)]\}^2,
\label{eq:edel90t}\\
M&=& {1 \over \bar{H}_c+16\bar{K}_{2}^{'}}.
\end{eqnarray}
Since $|\phi_{cl}(\taubar)| \ll 1$, we can obtain a reduced
one-dimensional action
by performing the 
Gaussian integration over $\phi$. In this case
Eq. (\ref{eq:edel90t}) is the effective potential energy in the
equation (\ref{eq:deltaeq})
for the bounce
$\delta_{cl}(\taubar)$.
Using Eqs. (\ref{eq:del90t}) and (\ref{eq:phi90t}),
the corresponding
classical action is given by
\ben
S_{cl}=(\hbar J){4\over 3}\sqrt{2}\epsn^{3/2}
(1-4{{K_2}\over K_1}).
\label{eq:scl90t}
\een
From Eq. (\ref{eq:edel90t}) and Fig.~2,
the height of barrier and the barrier
frequency for $\thH=\pi/2$ are given by
\begin{eqnarray}
 \bar{E}_1(\delm)=
 2\epsn^2[1-4(\kkbar)] \ \ \
 {\rm and}\ \ \  \ombbar=\sqrt{2} \bar{\omega}_c,
\end{eqnarray}
where $\delm=\sqrt{2\epsn}[1-4(\kkbar)]$.
Then the approximate form of the WKB exponent becomes
\ben
 B\sim {U\over \hbar \omega_b}=2J\epsn^{3/2}(1-4{K_2\over K_1}),
\een
which up to 
the numerical factor
is consistent with the action Eq. (\ref{eq:scl90t})
obtained by
the calculation with the explicit bounce solution.
\\

\subsection{$\thH \ll \thH \ll \pi$}

In this case
the potential energy and
the effective mass in the reduced one
dimension are approximately given by
\begin{eqnarray}
 \bar{E}_1(\delta)&=&-{1\over 2}[\bar{H}_c \sin(\thc-\thH)
                +2(\kkbar)\sin(4\thc)](\sqrt{6\epsn}
                \delta^2-\delta^3),
 \label{eq:edel135t}\\
 M&=&(1+|\cot\thH|^{2/3})
 [1-8{K_{2}^{'}\over K_1}+{2\over 3}
 ({3-2|\cot\thH|^{2/3}\over 1+|\cot\thH|^{2/3}})
 ({K_2- K_{2}^{'} \over K_1})].
\end{eqnarray}
The classical trajectory is given by
\begin{eqnarray}
 &\delta_{cl}(\taubar)& = \frac{\sqrt{6 \epsilon}}
 {\cosh^2(\bar{\omega}_t \bar{\tau})}, \\
 &\phi_{cl}(\taubar)& = -i (6\epsilon)^{3/4}
 |\cot\thH|^{1/6}
 \frac{\sinh(\bar{\omega}_t \bar{\tau})}
 {\cosh^3(\bar{\omega}_t \bar{\tau})}  \nonumber \\
 & &\ \ \ \ \times \sqrt{1+|\cot\thH|^{2/3}}
 [1-4{K_{2}^{'}\over K_1}+{2\over 3}
 ({2-3|\cot\thH|^{2/3}\over 1+|\cot\thH|^{2/3}})
 ({K_2- K_{2}^{'} \over K_1})],
\end{eqnarray}
where the precession frequency
$\bar{\omega}_t$ in MQT is given by
\begin{equation}
 \bar{\omega}_t=
 (\frac{3}{8})^{1/4}\epsilon^{1/4}(\frac{|\cotH|^{1/6}}
 {1+|\cotH|^{2/3}})
 [1+4{K_{2}^{'} \over K_1}
 +{2\over 3}
 ({5-3|\cot\thH|^{2/3}\over 1+|\cot\thH|^{2/3}})
 ({K_2- K_{2}^{'} \over K_1})].
\end{equation}
The corresponding classical action becomes
\begin{equation}
 S_{cl}= (\hbar J) \frac{16 \times 6^{1/4}}{5}
 \epsilon^{5/4} |\cot \theta_H|^{1/6}
 [1-4{K_{2}^{'} \over K_1}
 +{2\over 3}
 ({2-|\cot\thH|^{2/3}\over 1+|\cot\thH|^{2/3}})
 ({K_2- K_{2}^{'} \over K_1})].
 \label{eq:scl135t}
\end{equation}
Using the barrier height $\bar{E}_1(\delta_m=2\sqrt{6\epsilon}/3)
=4a(6\epsilon)^{3/2}/27$ 
where
$a=-[\bar{H}_c \sin(\thc-\thH)+2(\kkbar)\sin(4\thc)]/2$,
and the oscillation frequency
$\bar{\omega}_b=2\bar{\omega}_t$, 
we obtain the approximate form of the WKB exponent as
\ben
 B\sim \frac{U}{\hbar \omega_b}=\frac{4\times 6^{1/4}}{9}
 J \epsilon^{5/4} |\cot \theta_H|^{1/6}
 [1-4{K_{2}^{'} \over K_1}
 +{2\over 3}
 ({2-|\cot\thH|^{2/3}\over 1+|\cot\thH|^{2/3}})
 ({K_2- K_{2}^{'} \over K_1})],
\end{equation}
which is consistent with Eq. (\ref{eq:scl135t})
up to the numerical factor.\\

\subsection{$\thH=\pi$}

For $\theta_H =\pi$,
the total energy is
written as
\ben
 \bar{E}(\delta,\phi)=\bar{K}_{2}^{'}
 [1-\cos(4\phi)]\delta^4
 +{1\over 2}\{\epsn \delta^2-{1\over 4}
 [1-8(\kkbar)\delta^4]\}.
\een
From the Euler-Lagrange equation (\ref{eq:el1}) and
(\ref{eq:el2})
for $\delta(\taubar)$
and $\phi(\taubar)$, 
the classical trajectory is given by\cite{kim}
\begin{eqnarray}
 \delta_{cl}(\taubar)&=& \sqrt{\frac{K_1 \epsn}
 {\tilde{K}+K_{2}^{'}
 \cosh(4\bar{\omega}_t\bar{\tau})}}, \label{eq:delcl180t}\\
 \phi_{cl}(\taubar)&=& -i \bar{\omega}_t\bar{\tau}+{n\over 2}\pi,
 \label{eq:phi180t}
\end{eqnarray}
where $n=0, 1, 2, 3$, $\tilde{K}=K_1/4-K_2$, and
$\bar{\omega}_t=\epsn$.
Eq. (\ref{eq:phi180t}) shows that in the present case
$|\phi| \ll 1$ is not valid and that we
cannot expand $\bar{E}(\delta,\phi)$ as powers of
$\phi$, which means that
we cannot reduce the problem to the
one-dimensional one like Eq. (\ref{eq:deltaeq}).
Thus the classical action should be obtained directly
from the solution (\ref{eq:delcl180t}) and (\ref{eq:phi180t})
of the Euler-Lagrange equation, which becomes
\begin{equation}
 S_{cl}=(\hbar J)
 \frac{\epsn}{4}  (\frac{K_1}{K_{\alpha}})
 \ln ({\tilde{K}+K_{\alpha}\over K_{2}^{'}}),
\label{eq:scl180t}
\end{equation}
where $K_{\alpha}=\sqrt{\tilde{K}^2-(K_{2}^{'})^2}$.
Even though the effective potential energy and the mass
in one-dimensional form are not appropriate for the
present case, the dependence of $B$ on $\epsn$
can be derived from the values of $\bar{E}_1(\delta_m)$
and $\bar{\omega}_b$ in Table~I, which gives
$\epsn^{2-1}$ from $\bar{E}_1(\delta_m)(\propto \epsn^2)$
and $\bar{\omega}_b(\propto \epsn)$. \\

\subsection{Prefactor and Discussion}

Following the calculations used in the previous section,
we can
obtain the preexponential factor $C_0$ 
in Eq. (\ref{eq:rate}). As can be shown
in Table~II, for $\thH=\pi/2$ and $\pi/2 \ll \thH \ll \pi$
the values of $C_0$ in the tetragonal symmetry coincide
with those in the biaxial symmetry. This situation is
understood by the fact that the prefactor $C_0$ is determined
only by the shape of the effective potential energy in
the reduced one dimension, and the shapes are the same 
for the two
symmetries as can be seen in Eqs. (\ref{eq:en90b}),
(\ref{eq:en135b}), (\ref{eq:edel90t}), and 
(\ref{eq:edel135t}). However, the situation is
different for $\thH=\pi$ because $|\phi| \ll 1$ is not 
valid. Thus, the factor $C_0$ should be 
determined from $\delta^2 S$ by 
using the approximate forms of $\delta_{cl}$,
$E_{\delta \phi}, E_{\phi \phi}, E_{\delta \delta}$,
and so on, in the tetragonal 
symmetry case with $\thH=\pi$, which leads to
the result of $C_0$ in Table~II(b). 
It is possible to 
perform the same numerical calculation for 
given values of the 
parameters $\bar{K}_2$ and $\bar{K}_{2}^{'}$
in
the tetragonal symmetry in the same way as we did
in the biaxial 
symmetry for the full range
$\pi/2 < \thH \leq \pi$.
The results of the detailed 
numerical calculation for the
dependence of $B$ on $\thH$ will be
presented elsewhere.\\

\section{Conclusions}
\label{sec:conc}

In summary we obtained the approximate
analytic forms of the oscillation rate (MQC) for
$\thH=\pi/2$ and of the 
tunneling rate (MQT) for 
$\pi/2 \ll \thH \ll \pi$, and $\thH=\pi$
in the 
biaxial
and tetragonal symmetries.
Also, the $\thH$-dependence of
the WKB exponent $B$ for the full range 
$ \pi/2 < \thH \leq \pi$ with the values of
$\epsn=0.01$ and 0.001 was
calculated numerically in the biaxial symmetry. 
For future works 
it will be needed to compare
the theoretical results of the $\thH$-dependence
of $B$ for  given $\epsn$ and those of
the $\epsn$-dependence of $B$ for
given $\thH$ obtained in this paper
for the biaxial and teteragonal symmetries, with the 
experimental results which can be 
obtained by the same procedures as those suggested in 
Ref. \cite{mig} for the uniaxial symmetry.\\ 
 
\acknowledgments

This work was supported
in part by the Basic Science Research Institute Program,
Ministry of Education, Project No. BSRI-95-2414,
in part by the Ministry of Science and Technology of Korea
through the HTSRA, 
and in part by Non-Directed-Research-Fund,
Korea Research Foundation 1996.\\

\pagebreak
\begin{table}
\caption{ 
         The values appearing in Fig~2 for three
         ranges of $\thH$ (a) in the biaxial symmetry and (b) in
         the tetragonal symmetry. $\delta_m$ is the angle where
         the maximum of the barrier locates. Here,
         $a= -[\bar{H}_c \sin(\thc-\thH)+2\bar{K}\sin(4\thc)]/2$
         where $\bar{K}\equiv (K_2-K_{2}^{'})/2K_1$.}
 \label{tab:1}
\end{table}

\begin{center}
\begin{tabular}{|c|c|c|c|}  \hline
Field angle & $\delta_m$ & $\bar{E}_1(\delta_m)$ &
            $\bar{\omega}_b$  \\ \hline
$\thH=\pi/2$   &$\sqrt{2\epsn}$ &$\epsn^2/2$ 
&$\sqrt{2}\bar{\omega}_c$ \\
$\pi/2 \ll \thH \ll \pi$    &$2\sqrt{6\epsn}/3$
&$\sin(2\thc)(6\epsn)^{3/2}/27$ &2$\bar{\omega}_t$ \\
$\thH=\pi$    &$\sqrt{2\epsn}$ &$\epsn^2/2$
&$\sqrt{2}\bar{\omega}_{t}$ \\
\hline
\end{tabular}
\\
(a)
\end{center}

\vspace*{+1.0cm}
\begin{center}
\begin{tabular}{|c|c|c|c|}  \hline
Field angle & $\delta_m$ & $\bar{E}_1(\delta_m)$ &
            $\bar{\omega}_b$  \\ \hline
$\thH=\pi/2$   &$\sqrt{2\epsn}(1-4\bar{K})$ 
&$2\epsn^2 (1-4\bar{K})$ &$\sqrt{2}\bar{\omega}_c$ \\
$\pi/2 \ll \thH \ll \pi$    &$2\sqrt{6\epsn}/3$
&$4a(6\epsn)^{3/2}/27$ &2$\bar{\omega}_t$ \\
$\thH=\pi$    &$\sqrt{K_1 \epsn /2(\tilde{K}+K_{2}^{'}})$
&$K_1 \epsn^2 /8(\tilde{K}+K_{2}^{'})$ 
&$\sim \bar{\omega}_t$ \\
\hline
\end{tabular}
\\
(b)
\end{center}

\pagebreak
\begin{table}
\caption{The oscillation rate in MQC for $\theta_H=\pi/2$ and
         the tunneling rate in MQT for
         $\pi/2 \ll \thH \ll \pi$ and $\pi$ 
         (a) in the biaxial symmetry and 
         (b) in the tetragonal symmetry.
         $C_0$ is preexponential factor,
         $\omega_b$ in MQC or $\omega_p$ in MQT 
         the frequency of small oscillations around
         the minumum of the inverted potential, and
         $B$ the WKB exponent.
         The complete form of the rate is given in 
         Eq. (\protect\ref{eq:rate}). We define
         $K_{\beta} = (\tilde{K}+K_{\alpha})/{K_{2}^{'}}$,
         $K_{\gamma} = (K_2-K_{2}^{'})/K_1$,
         $\bar{\omega}_{u} = (\frac{3}{8})^{1/4}\frac{|\cotH|^{1/6}}
          {(1+|\cotH|^{2/3})}\epsilon^{1/4}$ and 
         $B_{u} = \frac{16 \times 6^{1/4}}{5} 
          |\cot\thH|^{1/6}\epsilon^{5/4}$,
         where $\bar{\omega}_{u}$ and 
         $B_{u}$ are the dimensionless frequency and
         the WKB exponent for $\pi/2 \ll \thH \ll \pi$
         in the uniaxial symmetry.\protect\cite{mig}}
 \label{tab:2}
\end{table}

\begin{center}
\begin{tabular}{|c|c|c|c|}  \hline
Field angle & $C_0$ & $\omega_I/\omega_0$ 
($I=c$ for MQC or $t$ for MQT) 
            & $B(=S_{cl}/\hbar)/J$ \\ \hline
$\thH=\pi/2$   &$8\sqrt{3}$ 
&$\frac{1}{\sqrt{2}}\sqrt{1+K_2/K_1}\epsilon^{1/2}$
&$\frac{4}{3}\sqrt{2}
(1+K_2/K_1)^{-1/2}\epsn^{3/2}$\\
$\pi/2 \ll \thH \ll \pi$    &$4\sqrt{15}$
&$\bar{\omega}_{u} [1+{K_2 \over K_1}(1+|\cot\thH|^{2/3})]^{1/2}$
&$\bar{B}_{u}/[1+{K_2 \over K_1}(1+|\cot\thH|^{2/3})]^{1/2}$\\
$\thH=\pi$    &$4\sqrt{3}$
&$\sqrt{\frac{K_2}{K_1}} \epsilon^{1/2}$
&$\frac{8}{3} \sqrt{\frac{K_1}{K_2}}\epsilon^{3/2}$\\
\hline
\end{tabular}
\\
(a)
\end{center}

\vspace*{1.0cm}
\begin{center}
\begin{tabular}{|c|c|c|c|}  \hline
Field angle & $C_0$ & $\omega_I/\omega_0$ (I=c or t)
            & $B(=S_{cl}/\hbar)/J$ \\ \hline
$\thH=\pi/2$   &$8\sqrt{3}$
&$\sqrt{\epsn \over 2}[1+2 (K_2+K_{2}^{'})/K_1]$
&$\frac{4}{3}\sqrt{2}
(1-4K_2/ K_1)\epsn^{3/2}$\\
$\pi/2 \ll \thH \ll \pi$    &$4\sqrt{15}$
&$\bar{\omega}_u
[1+4{K_{2}^{'} \over K_1}+{2\over 3}
({5-3|\cot\thH|^{2/3}\over 1+|\cot\thH|^{2/3}})
K_{\gamma}]$
&$\bar{B}_u
[1-4{K_{2}^{'} \over K_1}+{2\over 3}
({2-|\cot\thH|^{2/3}\over 1+|\cot\thH|^{2/3}})
K_{\gamma}]$\\
$\thH=\pi$    &$16\sqrt{K_{\alpha}\over K_{2}^{'}}
{ {K_{\beta}^{-\tilde{K}/2K_\alpha}} \over
{\sqrt{\ln K_\beta}} }$
&$\epsn$
&$\frac{1}{4} {K_1\over K_{\alpha}}
[\ln K_{\beta}] \epsn$\\
\hline
\end{tabular}
\\
(b)
\end{center}

\pagebreak

\begin{figure}
 \caption{Comparison of (a) $\sin (2 \theta_c)$ with
         (b) $\cos (2 \theta_c)$ in 
         Eqs. (\protect\ref{eq:etaeq})
         and (\protect\ref{eq:edelta}).
         Note that $\sin (2 \theta_c)=0$,
         $|\cos (2 \theta_c)|=1$ for both
         $\theta_H = \pi/2$ and $\pi$.}
 \label{fig:1}
\end{figure}

\begin{figure}
 \caption{The plots of the effective potential  
         $\bar{E}_1(\delta)$ as a function of
         $\delta(=\theta-\thz)$ for 
         (a) $\thH=\pi/2$ (MQC), (b) $3\pi/4$ and (c) $\pi$
         about $\thz$ at which 
         the system is metastable (MQT).
         The position $\delta_m$ 
         and the height $\bar{E}_1(\delta_m)$ of the
         barrier, and the barrier
         frequency $\bar{\omega}_b$ are given in Table~I for 
         the biaxial
         and tetragoanl symmetries.}
 \label{fig:2}
\end{figure}

\begin{figure}
 \caption{The $\thH$-dependence of the instanton solutions in
          the biaxial symmetry for $\epsn(=1-H/H_c)=0.001$ 
          and $K_1=K_2$.
          Here, 
          (a) $\thH = 90.0002^{\circ}$, (b) $90.01^{\circ}$,
          (c) $135^{\circ}$, (d) $179.99^{\circ}$,
          (e) $180^{\circ}$.}
 \label{fig:3}
\end{figure}

\begin{figure}
 \caption{The $\thH$-dependence of the 
          relative WKB exponent $B(\thH)/B(3\pi/4)$ 
          in the biaxial symmetry with $K_1=K_2$ for
          (a) $\epsn=0.01$ and (b) $\epsn=0.001$. Here,
          $B(\thH=3\pi/4,\epsn=0.01)=9.14\times 10^{-3}$
          and 
          $B(\thH=3\pi/4,\epsn=0.001)=5.14\times 10^{-4}$.}
 \label{fig:4}
\end{figure}


\begin{references}
\bibitem{foot0} We have also performed the calculations for 
the hexagonal
symmetry with the easy direction along the c axis and 
for the cubic symmetry
with $K_1 > 0$ and $K_1 < 0$. 
In case of the hexagonal symmetry the approximate 
formulas are very complicated, and we do not present them here.
We just note that as the driving term is 
$K_{2}^{'}$ term in the tetragonal symmetry, it is 
the $K_{3}^{'}$ term in the hexagonal symmetry, which
leads to the results of $\Gamma, \bar{\omega}_t,
\bar{H}_c, \thc$, etc., which also depend on $K_2$ and
$(K_3-K_{3}^{'})/K_1$.
In case
of the cubic symmetry, the easy axes are [100], [001], etc., for 
$K_1>0$ and [111], [$\bar{1}\bar{1}$1], etc., for $K_1<0$.
Thus the range of $\thH$ in which we are interested is
$\pi/4 \leq \thH \leq \pi/2$ for $K_1>0$ 
and $0 \leq \thH \leq \sin^{-1}(1/\sqrt{3})$ for $K_1<0$.
Even though the anisotropy energy for $K_1>0$ can be
reduced to the one
for the tetragonal symmetry by replacing $\bar{K}_2$ and 
$\bar{K}_{2}^{'}$ by $-7/16$ and $1/16$, the approximation which we 
used in the tetragonal symmetry is not applicable. 
Also, the situation is much more complicated in case of the cubic
symmetry for $K_1<0$.
For the cubic symmetry, numerical approach should be used
to obtain the dependence of tunneling rate on the 
direction of magnetic field, which will be done elsewhere.
\bibitem{com} In fact, for 
$\pi/2 \ll \thH \ll \pi$
our result of $C_0$ is different from that in
Ref. \cite{mig} by a factor
of 4. We carefully checked that our results is 
consistent with that of 
Ref. \cite{cal} which is referred in Ref. \cite{mig}.
\bibitem{foot1} In Ref. \cite{mig} it is discussed that  
the sharpness of the action
for $\thH=\pi$ comes from the behavior of $|\cotH|$
in the uniaxial symmetry. 
As can be seen from Table~I, we notice that actually 
the sharpness of the WKB exponent
is not from the $|\cotH|$ term, because 
there is difference in the $\epsn$ dependences for
$\thH=\pi$ and for $\pi/2 \ll \thH \ll \pi$,
which comes from
$\sin(2\thc)$ or $\cos(2\thc)$ term 
in Eq. (\ref{eq:edelta}).
\bibitem{foot2}
In this work, the approximate form of the physical quantities 
in the tetragonal symmetry are obtained for
$|\kkbar| \ll 1$. Numerical calculation should
be performed in other cases.
\end{references}
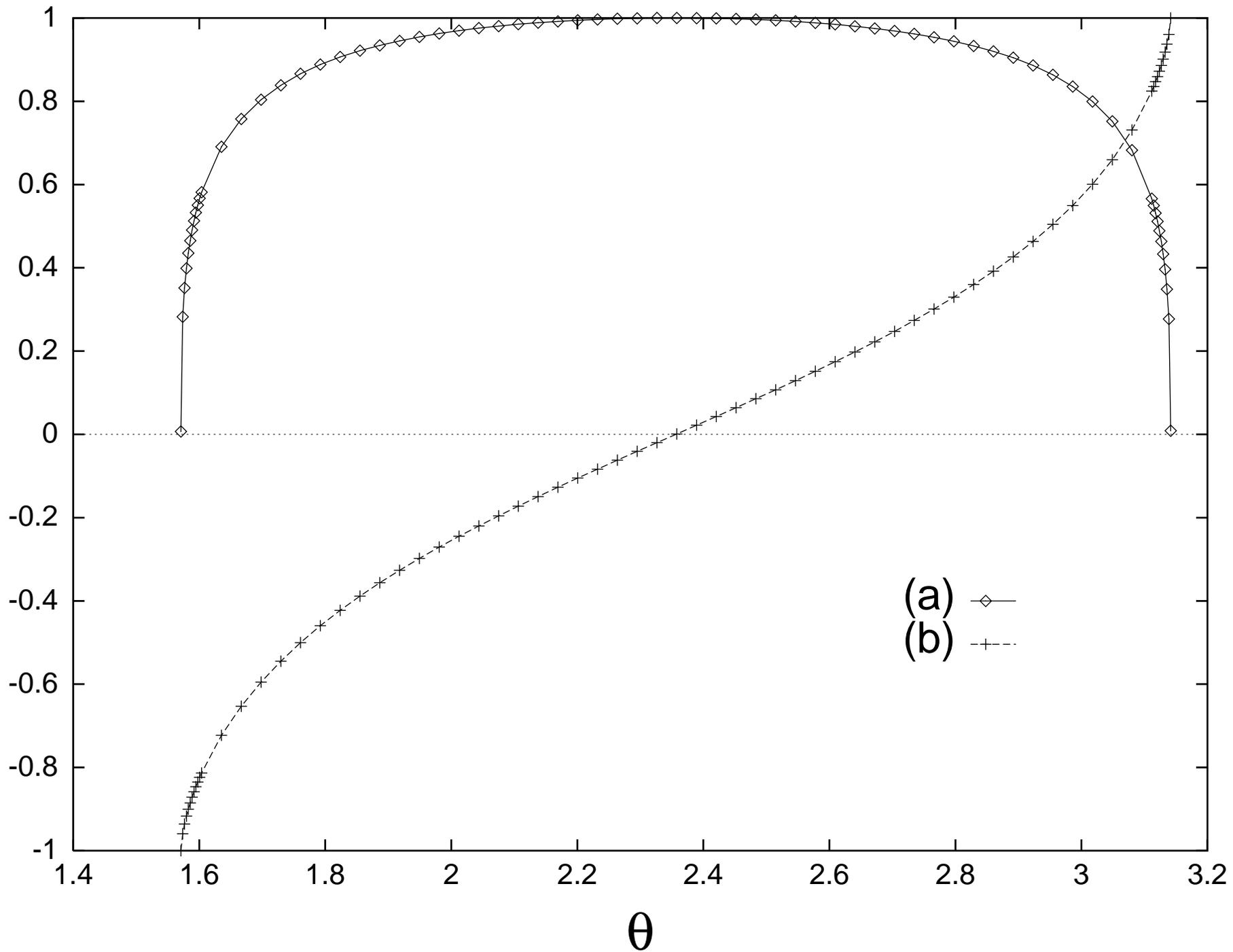
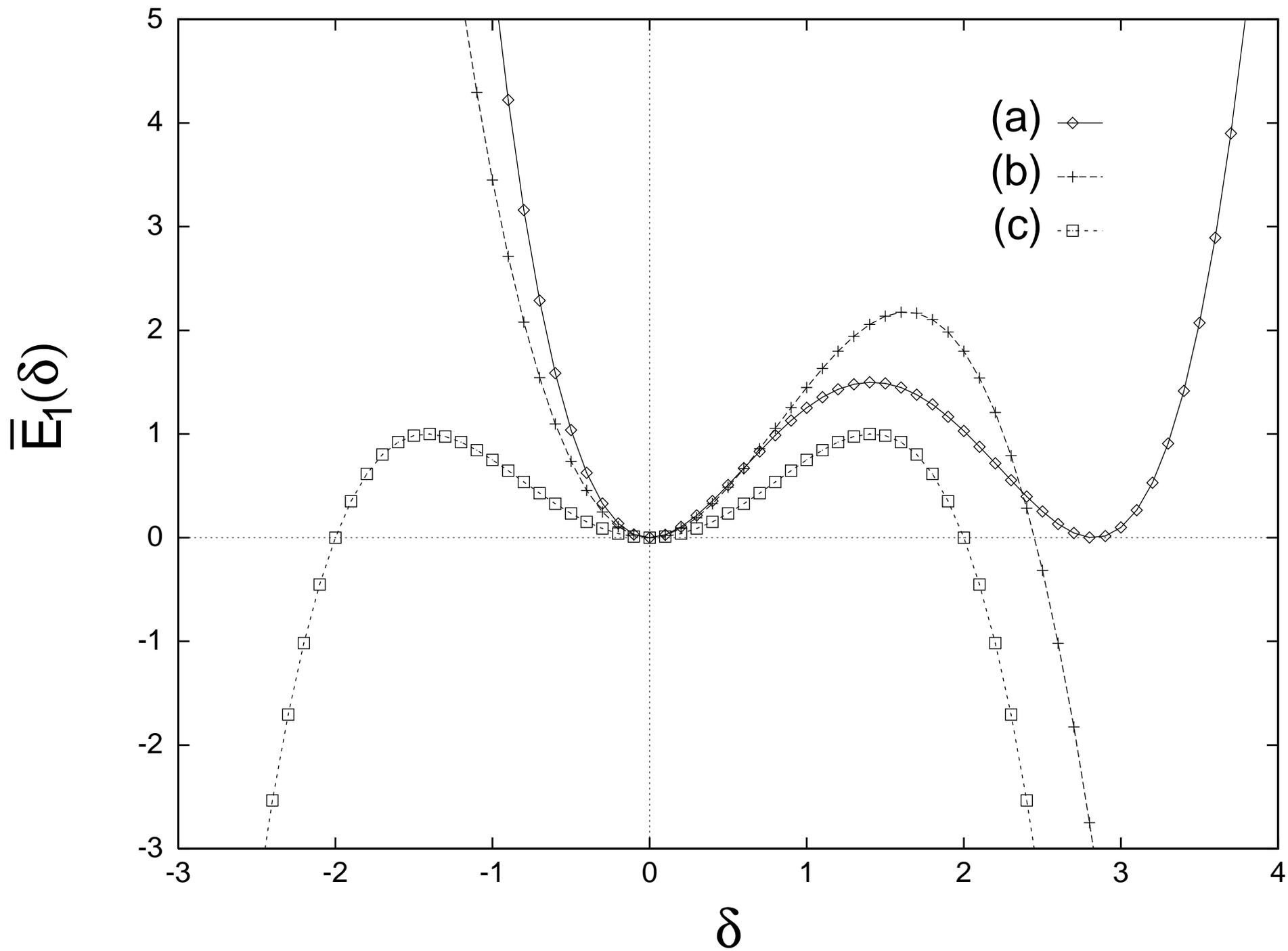
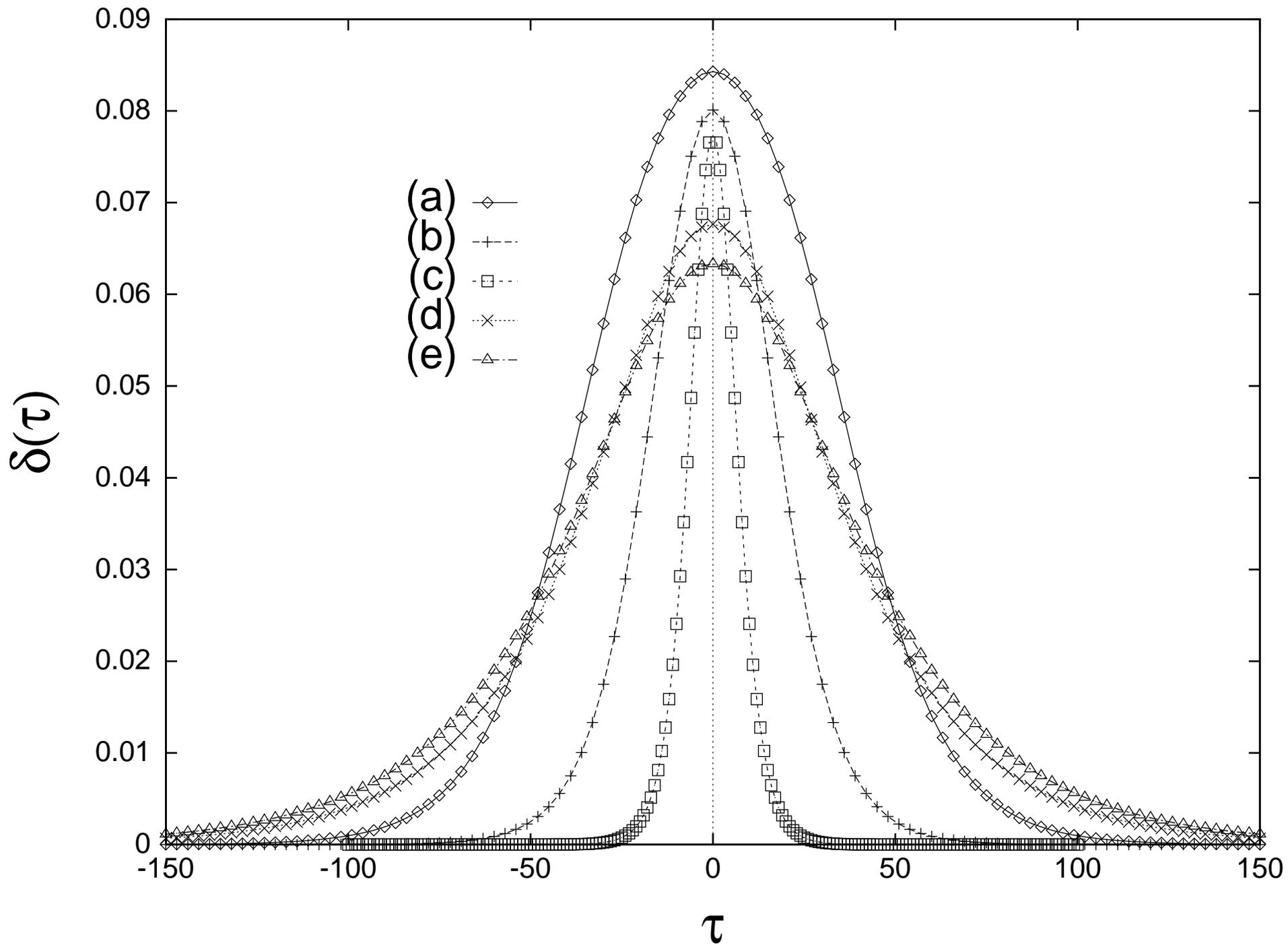
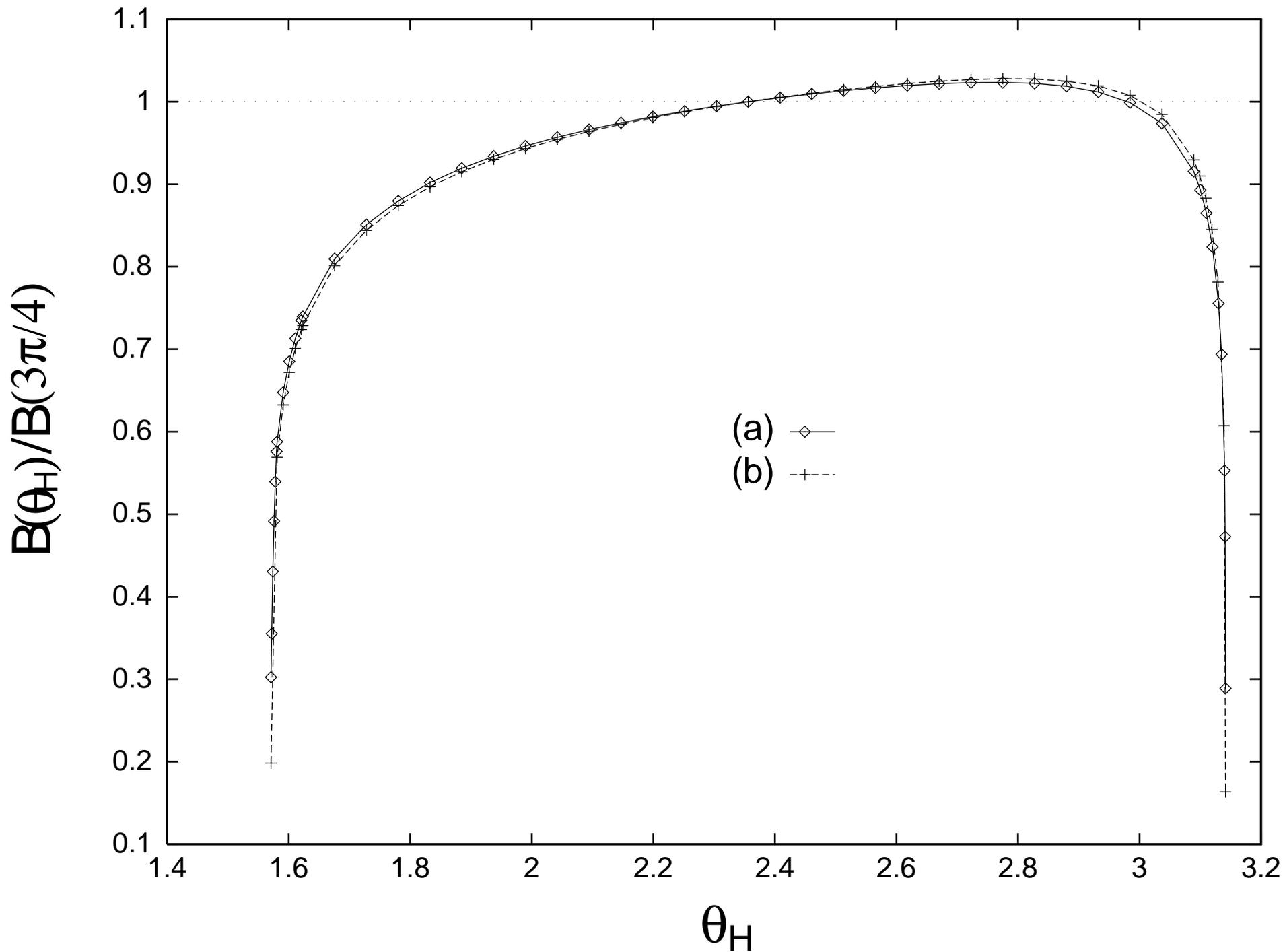
\end{document}